\documentclass{ws-ijmpcs}

\begin{document}

\markboth{Manel Perucho}
{Jet propagation and deceleration}

%
\catchline{}{}{}{}{}
%

\title{Jet propagation and deceleration}

\author{Manel Perucho}

\address{Departament d'Astronomia i Astrof\'{\i}sica, Universitat de Val\`encia, C/ Dr. Moliner, 50\\
Burjassot, Valencian Country 46100, Spain\\
manel.perucho@uv.es}
\maketitle

\begin{history}
\received{Day Month Year}
\revised{Day Month Year}
\end{history}

\begin{abstract}
 Extragalactic jets in active galactic nuclei (AGN) are divided into two morphological types, namely Fanaroff-Riley I (FRI) and Fanaroff-Riley II (FRII). The former show decollimated structure at the kiloparsec scales and are thought to be decelerated by entrainment within the first kiloparsecs of evolution inside the host galaxy. The entrainment and deceleration can be, at least partly, due to the interaction of jets with stellar winds and gas clouds that enter in the jet as they orbit around the galactic centre. In this contribution, I review recent simulations to study the dynamic effect of entrainment from stellar winds in jets and the direct interaction of jets with gas clouds and stellar winds. I also briefly describe the importance of these interactions as a possible scenario of high-energy emission from extragalactic jets. 

\end{abstract}

\ccode{PACS numbers: 47.75.+f, 98.62.Nx, 98.54.-h}

\section{Introduction}	

The well-known morphological dichotomy between FRI and FRII jets\cite{fr74} from AGN has been claimed to be related with jet power.\cite{rs91} Within this picture, on the one hand, the extragalactic FRII jets show an edge-brightened structure in the radio correspond to those with powers typically over $10^{44}\,{\rm erg/s}$. In this case, the jets show a remarkable stability and reach the interaction site with the ambient still collimated, generating a strong forward bow-shock, and a reverse shock, known as hot-spot. On the other hand, the edge-darkened jets in radio are related with lower powers and, although they show a collimated morphology in the parsec scales, there is a transition to large opening angles or decollimation along the first kiloparsecs that leads to the absence of a hot-spot at the interaction regions. The evidence for deceleration in the latter comes from the symmetry of the jet/counter-jet emission at those large scales, as opposed to the asymmetries observed in the case of FRII jets due to Doppler boosting of the radiation in the direction of the propagation of the flow. Of course, jet power by itself is not an explanation for jet deceleration and the suggested reason for deceleration has been typically attached to entrainment processes of cold, slow gas that is incorporated and mixed with the jet flow.\cite{bi84}\cdash\cite{la96} 

In Ref.~\refcite{bi84}, the author suggested that turbulent mixing via the development of a shear-layer could explain the deceleration of FRI jets at kiloparsec-scales. Later, in Refs.~\refcite{lb02} and \refcite{wa09} detailed models were developed, both based on this idea. Numerical simulations of the development of Kelvin-Helmholtz instabilities in two and three dimensions also support this idea.\cite{pe04}\cdash\cite{pe10} Another possible, compatible, scenario is the interaction of jets and clouds or stars, initially proposed by Ref.~\refcite{bk79} to explain the presence of knots in M87 (see also the numerical simulations in Ref.~\refcite{cb85}), and later studied by Refs.~\refcite{ko94}, \refcite{bo96}, and \refcite{hb06} as a possible source of mass-load to the jet flow. The main conclusions of these works are the following: 1) The mass-load by stellar winds can be treated as a hydrodynamical problem\cite{ko94}\cdash\cite{bo96}; 2) the mass load of an old population of stars, as that expected in giant elliptical galaxies typically hosting jets, can efficiently decelerate low power jets, i.e., FRIs,\cite{ko94}\cdash\cite{bo96} and 3) A single star with large mass-loss rates, e.g., a Wolf-Rayet, can temporally quench the whole jet flow for jet powers $L_j\leq 10^{42}\,{\rm erg/s}$.\cite{hb06} From an observational perspective, mass-loading by stellar winds has been recently claimed to explain the pressure imbalance between the lobes and the ambient medium in the lobes of Centaurus~A.\cite{wy13} Thus, there is consensus that entrainment by stellar winds could be dynamically relevant for jet evolution. It has also been suggested\cite{pm07} that, in the case of the higher power FRI jets, a strong recollimation shock could be responsible for the triggering of instabilities that produce shear-layer mixing and deceleration. 

In addition, the radiative counterpart of the interaction of relativistic flows with stellar winds or clouds can also be important, due to the strong shock produced\cite{bp97}\cdash\cite{ba12b} and the subsequent particle acceleration, which could be responsible for the production of the variable gamma-ray emission observed in a number of AGN jets. Finally, recent works about the knots observed in the jet in Centaurus~A suggest that, tens to hundreds of parsecs away from the source, those knots could correspond to this kind of interactions.\cite{wo08}\cdash\cite{go10}  

During the last years we have performed numerical simulations of the influence of mass-loading by stellar winds on low-power jets,\cite{pe13} and on the detailed process of jet/cloud\cite{brp12} and jet/stellar wind interaction. In this contribution, I review these recent results on the study of the effect that mass-load by stars and clouds may have on jet dynamics, and the importance of these processes for high-energy radiation. 
 The paper is structured as follows: Section~\ref{s:sim} is focused on the setup of the simulations performed for this study. Section~\ref{s:res} is devoted to the results and their importance to high-energy radiation, and the main results and conclusions of this work are given in Section~\ref{s:sum}.
 
\section{Simulations}\label{s:sim}
  In this section the basics of the initial setups for the simulations are given. Only the main physical characteristics are explained. A detailed description of these simulations can be found in Ref.~\refcite{brp12}, and Perucho et al. (in preparation).   
\subsection{Long-term evolution}
  We have performed jet evolution simulations\cite{pe13} in which different jets are injected into an ambient medium composed of a decreasing
density atmosphere of neutral hydrogen in hydrostatic equilibrium as that given in Ref.\refcite{pm07}. In addition, we have
included a law that accounts for mass entrainment into the jets from stellar-winds obtained from
a Nuker distribution of surface brightness in an elliptical galaxy.\cite{la07} This is done by assuming that the stellar population in a typical giant elliptical jet-hosting galaxy is basically composed of low-mass stars, with stellar winds of $10^{-11}-10^{-12}\,{\rm M_{\odot} yr^{-1}}$. The mass-load is treated as a source term in the mass equation. The grid extends from 80~pc to 2~kpc with a resolution of 16 cells per beam radius at injection ($R_j=10\,{\rm pc}$). The simulations are two-dimensional and axisymmetric. 
  Regarding the injected jets, they are injected as purely electron/positron outflows with powers that range between $5\times10^{41}$ and $10^{44}\,{\rm erg/s}$, density ratios ranging from $10^{-5}$ to $10^{-10}$ times the ambient density at injection $\rho_{a,80}=3\times 10^{-25}\,{\rm g/cm^{3}}$, temperatures between $10^9$ and $3\times10^{13}~{\rm K}$, and injection velocities of $0.95\,c$ in the case of the lowest power jets, and $0.99\,c$ in the case of the most powerful ones. One of the low-power jet simulations was repeated without the mass injection from the winds to check the different behavior. 
\subsection{Jet-cloud/star interaction}
   We have performed a second set of simulations consisting of a cloud (CJ simulations, from now on) or a spherically symmetric stellar wind (injected from a number of cells, SW simulations, from now on) located on the symmetry axis of a two-dimensional axisymmetric grid, and interacting with a jet flow, which fills the rest of the numerical grid.\cite{brp12} In CJ simulations, the jet parameters are taken to give a total jet power $\simeq10^{44}\,{\rm erg/s}$ for jet radius $R_j=10^{15}\,{\rm cm}$, this is, in compact jet regions, within the broad-line region (BLR) of the AGN. In SW simulations the jet power is kept similar to CJ simulations, but the jet radius is taken to be $R_j=3.1\times10^{19}\,{\rm cm}\,=\,10\,{\rm pc}$, because SW simulations are aimed to study the interaction of the jet with a red giant at 100~pc from the central black-hole.\cite{pe14}     
   In addition, we currently perform a three-dimensional simulation of the process of entrance of a red-giant stellar wind bubble into a jet. In this case, a part of the grid is a steady ambient and another part is filled by a relativistic flow in pressure equilibrium with the ambient. The injector of the stellar wind in the numerical grid is initially located within the ambient and it propagates towards the jet with a typical velocity around the galactic centre.\cite{pe14}

\section{Results}\label{s:res}
\subsection{Long-term evolution}
\begin{figure}[pt]
\centerline{\psfig{file=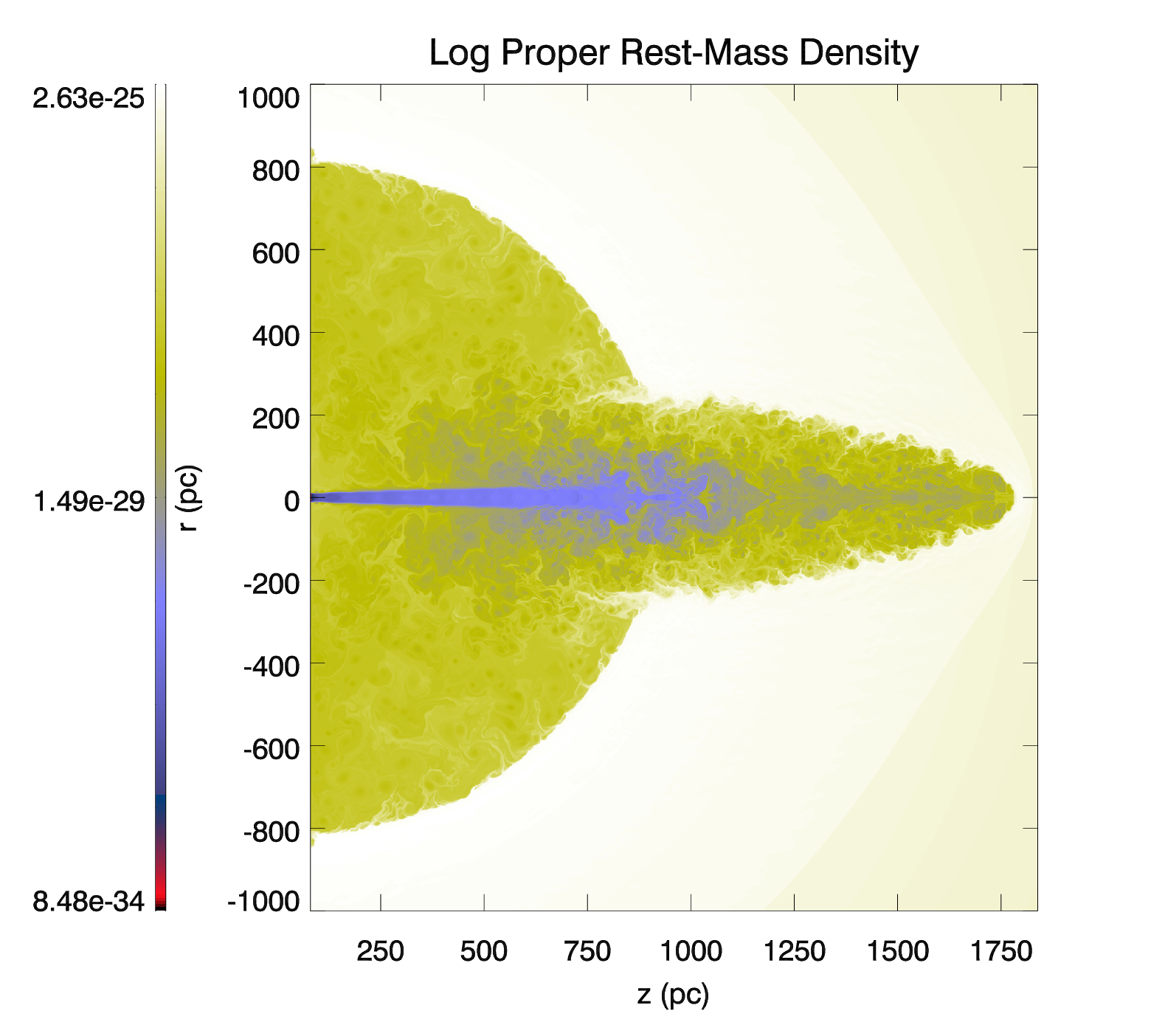,width=0.8\textwidth}}
\centerline{\psfig{file=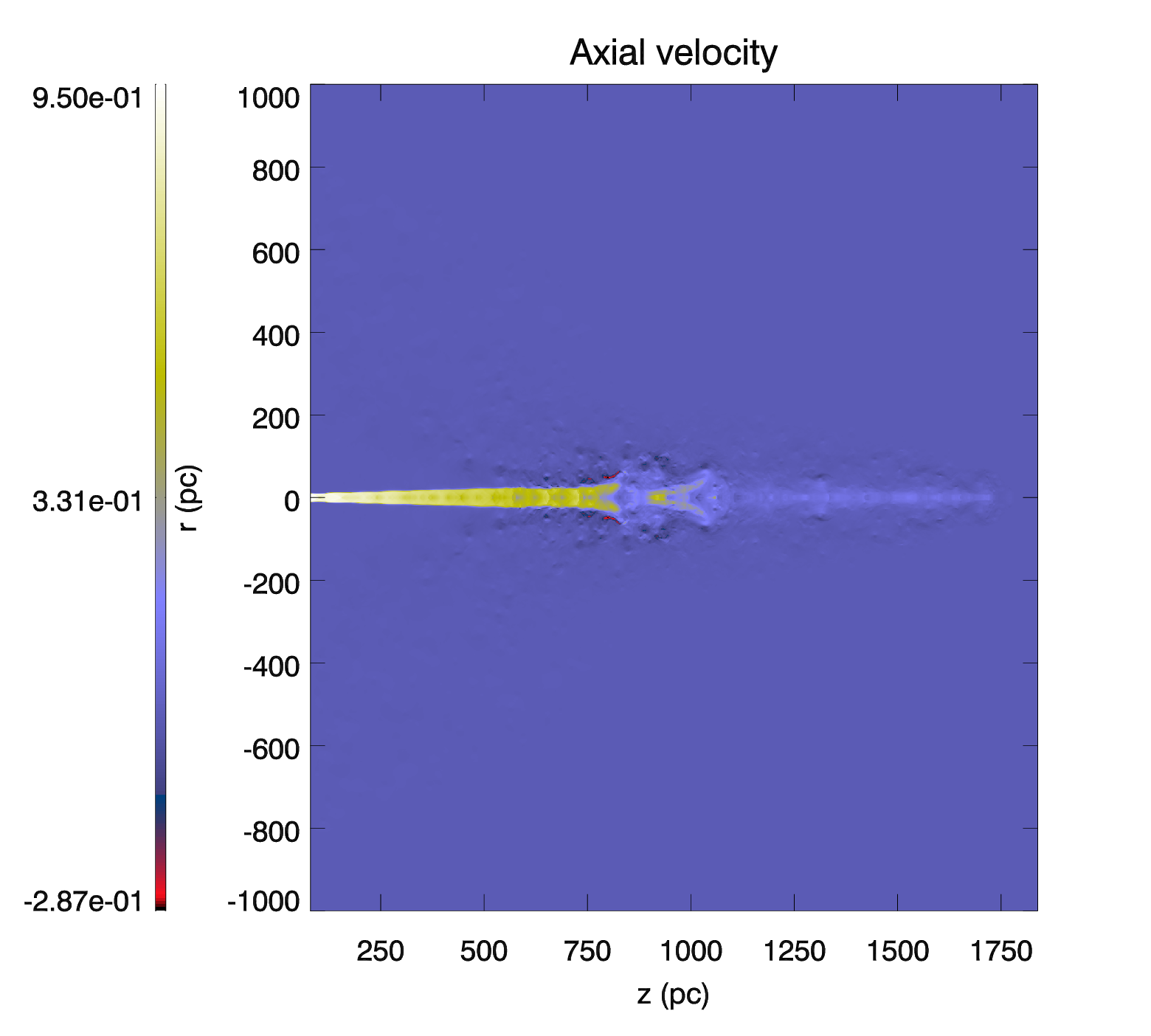,width=0.8\textwidth}}
\vspace*{8pt}
\caption{Rest-mass density (${\rm g/cm^{3}}$, top) and axial velocity (in units of the speed of light, $c$) of a simulated jet with $L_j=5\times10^{41}\,{\rm erg/s}$, at $t\simeq2\times10^6\,{\rm yrs}$. The simulation is axisymmetric. \label{f1}}
\end{figure}

 The simulations of jet evolution point towards mass-load by stellar winds being unable to stop the most powerful FRI jets ($L_j\simeq 10^{44}\,{\rm erg/s}$) alone. However, in the case of low-power jets, the jets are clearly decelerated due to the mass-load. In addition, they expand and cool due to the entrainment of the colder stellar wind particles. Figure~\ref{f1} shows the rest mass density and axial velocity of one of these simulations. The jet density is clearly increasing and its velocity decreasing with distance. The increase in the jet radius is due to the increase in the pressure within the jet, which is induced by the injection of a significant number of particles together with the deceleration of the flow. In addition, even a small pressure mismatch with the ambient triggers the development of a pinching Kelvin-Helmholtz mode that grows with distance. The growth of this mode is larger when the Mach number of the flow decreases. Thus, at some point, the unstable mode becomes nonlinear and disrupts the jet flow. After disruption, a subrelativistic outflow remains. In the top panel of Fig.~\ref{f1}, the cocoon of the flow, which is composed of shocked jet material mixed with shocked ambient material through the contact discontinuity, shows a peculiar structure. This is due to the long dynamical time-scales, which allow the gravity that keeps the atmosphere at rest to act on this material and configure it in a spherical shape. 
  
  In the case of high-power FRI  jets, another mechanism, such as a strong recollimation shock\cite{pm07} should be responsible for the deceleration of the flow at the kiloparsec scales. It is important to remark that in these simulations we have considered a homogeneous population by small mass, old stars. However, if the stellar population of a galaxy is shifted towards stars with more powerful winds, this process could efficiently decelerate jets with powers $10^{43}\,{\rm erg/s}<L_j<10^{44}\,{\rm erg/s}$. Another important conclusion of these simulations is that low-power FRI jets cannot be magnetically dominated at kiloparsec scales as it has been suggested (see, e.g., Ref.~\refcite{nmg11}), due to the mass-load by the stars in the host galaxy, which may completely change the nature of the outflow within hundreds of parsecs.\cite{wy13}

\subsection{Jet-cloud/star interaction}
\begin{figure}[pt]
\centerline{\psfig{file=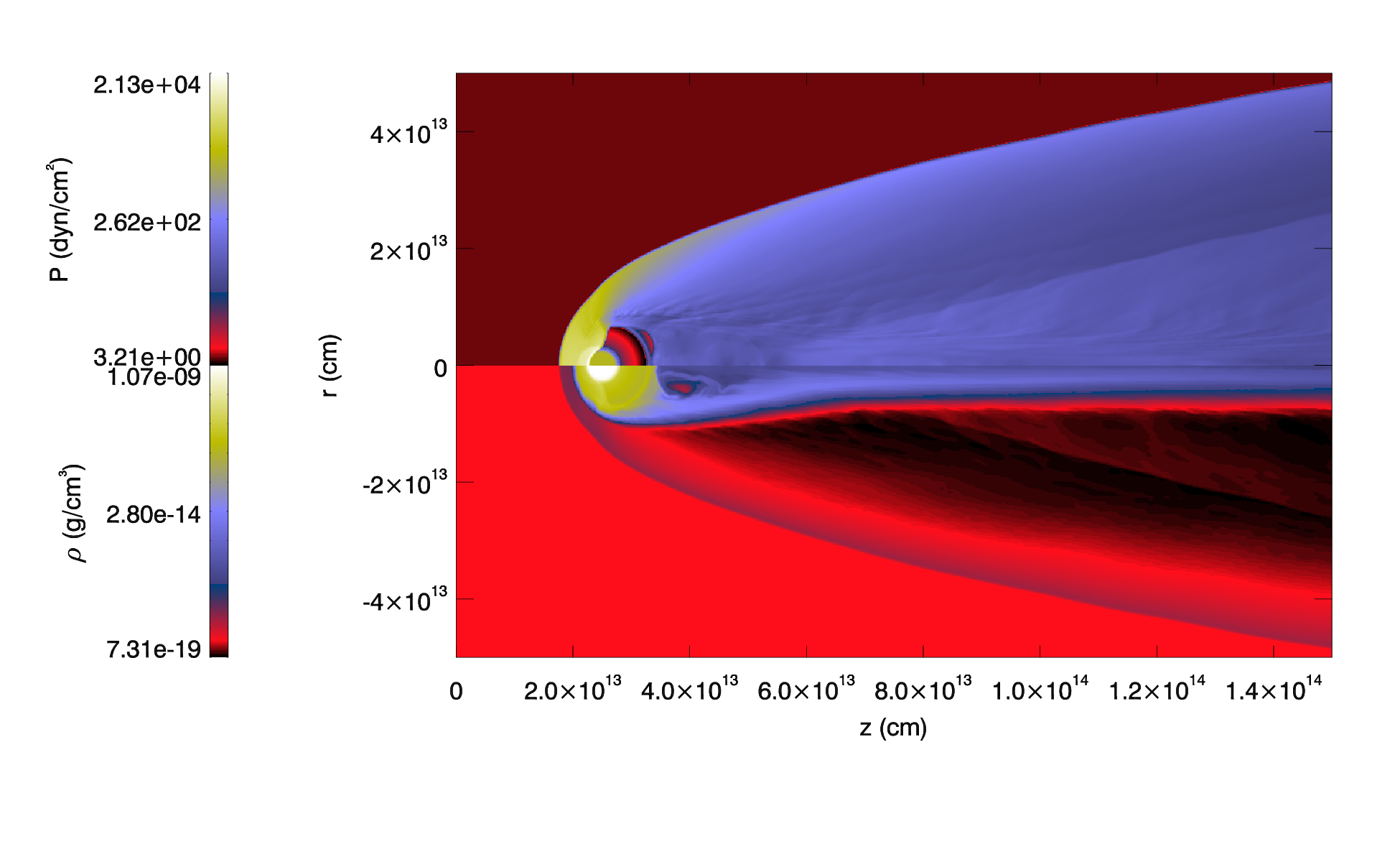,width=\textwidth}} 
\centerline{\psfig{file=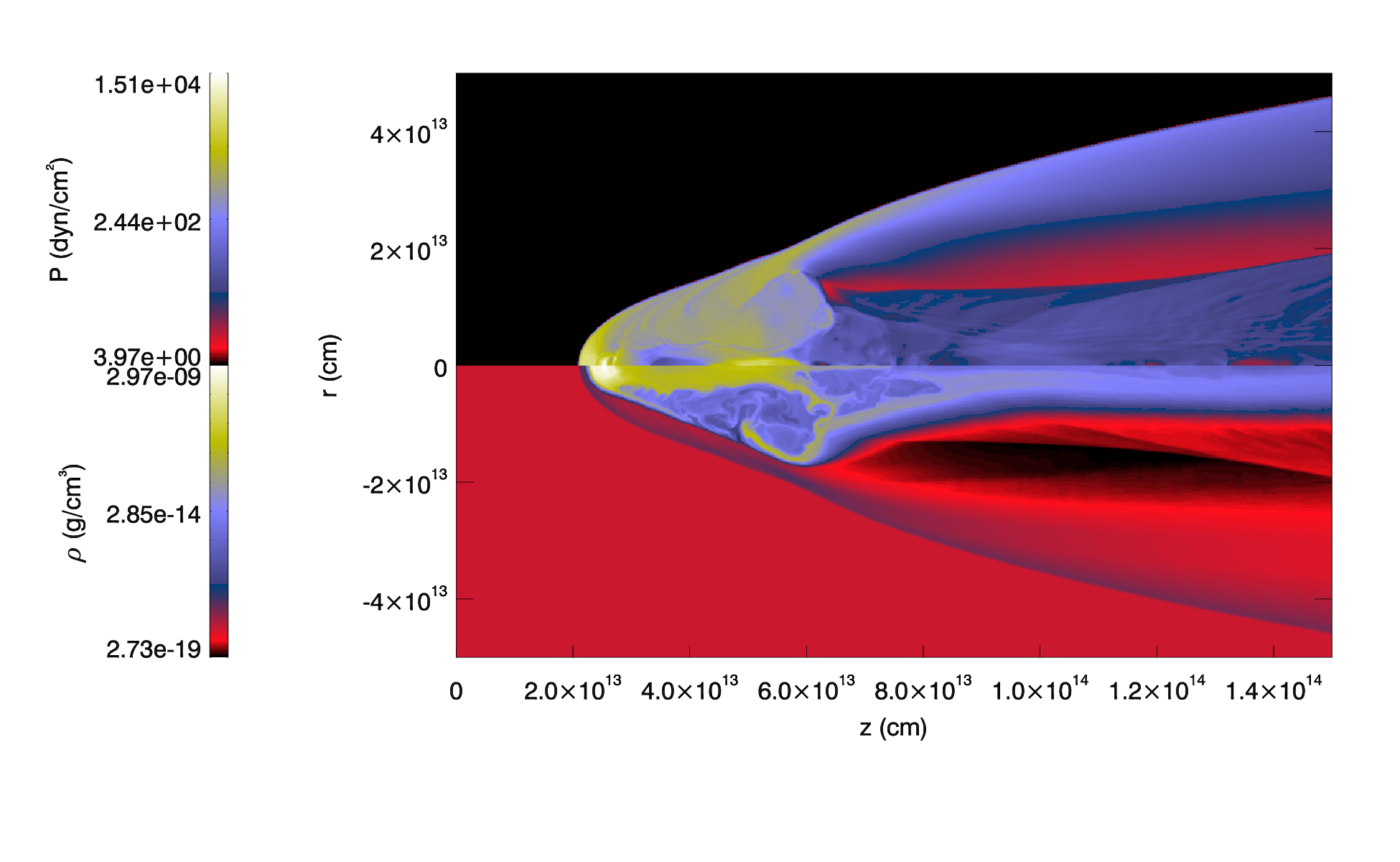,width=\textwidth}}
\vspace*{8pt}
\caption{Rest-mass density (${\rm g/cm^{3}}$, lower half of the panels) and pressure (${\rm dyn/cm^{2}}$, upper half of the panels) of an axisymmetric simulation of a jet interacting with an inhomogeneous cloud at the BLR of an active galaxy. The top panel shows one of the initial plots of the shock propagating through the diluter region of the cloud at time $1.6\times10^{5}\,{\rm s}$. The bottom panel shows a frame in which the diluter part of the cloud has been disrupted and a large portion of it is being accelerated downstream, at time $4.3\times10^{5}\,{\rm s}$. \label{f2}}
\end{figure}

CJ simulations focused on the interaction of a jet with a cloud within the BLR of an active galaxy. In the case of a homogeneous cloud, the impact of the jet flow with the cloud gas generates a shock wave that propagates through the cloud. The shock heats up the cloud and results into an expansion, increasing its cross section. Finally, when the shock has crossed the whole cloud, the cloud is disrupted into small pieces, which are again disrupted by the same mechanism, in a \emph{self-similar} way. In the case of an inhomogeneous cloud, with a denser core, the shock heats and disrupts the outer, more diluted layers of the clouds, but the dense core remains basically untouched during the whole simulation. Figure~\ref{f2} shows the rest-mass density and pressure structure of the jet-cloud interaction in the inhomogeneous case at two different times. The first image shows a frame in which the shock is propagating through the outer layers, whereas in the case of the second, the shock has completely crossed this region and all its remaining gas is being dragged and accelerated downstream.

In both cases, the shocked cloud material forms a cometary-like tail, in which the gas is accelerated downstream. The simulations focus on a very small region to study the interaction in detail, so at the end of the grid the tail configuration is still stable and we observe no mixing with the jet. However, already within the grid we observe acceleration of the tail flow due to expansion, and in addition, farther downstream, the growth of instabilities is expected, which must eventually end in mixing with the jet flow. 

\begin{figure}[pt]
\centerline{\psfig{file=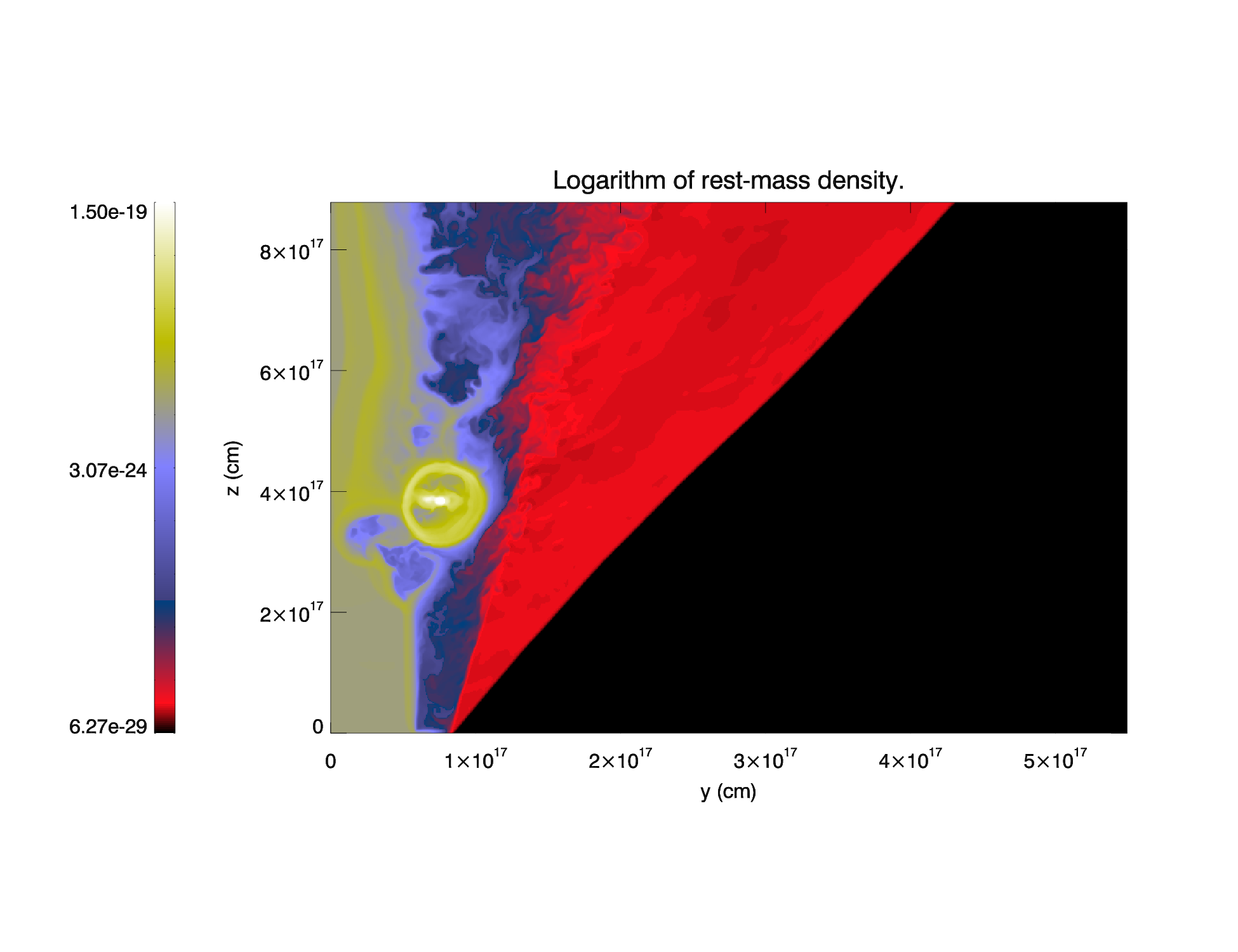,width=\textwidth}} 
\vspace*{8pt}
\caption{Rest-mass density (${\rm g/cm^{3}}$) cut of a 3D simulation of a red-giant like injection region penetrating into a relativistic flow. The low-density black region on the right of the image represents the jet, with the flow propagating in the upward direction. In the shear-layer, a wave propagates upstream due to the lower local axial speed. A strong shock propagates into the jet and decelerates the shocked jet flow. \label{f3}}
\end{figure}

To further investigate the process of mass-loading, we performed a two-dimensional axisymmetric simulation of a red giant wind within a jet, at 100~pc from the active nucleus. In this case, we added an extended grid along the axial direction, with geometrically increasing cell size, to follow the cometary tail formed by the shocked wind from 100~pc to 200~pc. In this case, however, the strong wind of the red giant gives rise to a high density tail that is very stable against the growth of pinching Kelvin-Helmholtz modes. In three-dimensional simulations we would expect the trigger of the faster growing helical modes that could favor efficient mixing with the jet flow. We currently perform a three-dimensional simulation of a similar red-giant wind injection region into a relativistic jet (see Fig.~\ref{f3}). A strong shock is formed at the entrance of this \emph{obstacle} into the jet. Our aim is to study the full process of entrance of the star into the jet and the resulting configuration once the star has fully penetrated the jet.\cite{pe14} 

\subsection{Importance for high-energy radiation}
The interaction of jets with clouds and/or stars, or the presence of strong recollimation shocks in jets, is not only important from a dynamical point of view, but also due to its radiative implications. Our simulations show that jet-cloud interaction can convert a significant fraction of the local jet's kinetic energy into thermal energy and potentially radiation, and this scenario has been claimed as a possible way to explain high-energy emission from compact jet regions, within the BLR of active galactic nuclei.\cite{bp97}\cdash\cite{ba12b} In Ref.~\refcite{brp12} we concluded that this claim is plausible. Regarding strong recollimation shocks, if a Mach disk is formed, a large amount of the jet energy flux, initially in the form of kinetic energy, can be converted into internal energy and an important number of particles can be accelerated, thus increasing the emissivity throughout the electromagnetic spectrum. Also the evolution simulations show that the kinetic energy of hot, diluted outflows can be efficiently converted into thermal energy. In this case, we simulated the entrainment by the stellar winds of low-mass stars. This interaction with low-mass stars is possibly unable to produce high-energy emission, due to the lower density of the stellar winds. However, the dissipation of the kinetic energy of the jet at multiple shocks plus the process of turbulent mixing within the jet could well be partly responsible for the multifrequency emission detected from FRI jets at kiloparsec scales, from radio to X-rays, as it is the case of Centaurus~A.\cite{wo08}\cdash\cite{go10} 

\section{Summary}\label{s:sum}
We currently perform a long-term project to study the mass-load and deceleration of low-power jets within the first kiloparsecs, and the radiative implications that this process may have. Our main conclusions regarding the role of mass-load by stellar winds in the deceleration of FRI jets are that: 1) the typical low-mass, old stellar populations of massive elliptical galaxies are only able to decelerate jets with powers $L_j\leq10^{42}\,{\rm erg/s}$,\cite{pe13} 2) in the case of large power FRI jets ($L_j\sim10^{44}\,{\rm erg/s}$), the recollimation shock scenario is favored.\cite{pm07,pe13} This shock should be spatially coincident with the beginning of the \emph{flaring region} observed in these jets. 

 Our main conclusions regarding the jet-cloud/star interaction are: 1) A cometary tail is formed as the shock heats, inflates and ablates the gas, 2) the gas is accelerated along the tail due to expansion, 3) depending on the properties of the cloud/wind material, it can be more (denser clouds/winds) or less (diluter clouds/winds) stable against the growth of instabilities, but in any case, the tail material will be completely mixed and incorporated within the jet flow via the growth of a turbulent boundary layer,\cite{ko94} 4) homogeneous clouds can be completely disrupted and advected with the jet, 5) a large fraction of the local jet energy flux can be converted into internal energy and particle acceleration can occur at this kind of strong interactions.
 
  Finally, the work done by us and other groups shows that the physical conditions produced at strong recollimation shocks or interaction shocks represent plausible scenarios to produce emission through the whole electromagnetic spectrum. Future work will include the extension of our study to 3D relativistic magnetohydrodynamical simulations.

\section*{Acknowledgments}
I thank my collaborators in all these works for interesting discussions and for sharing their knowledge with me.
I acknowledge financial support by the Spanish ``Ministerio de Ciencia e Innovaci\'on'' (MICINN) grants AYA2010-21322-C03-01, AYA2010-21097-C03-01 and CONSOLIDER2007-00050.


\end{document}